\input{aipcheck}

\documentclass[
    ,final            
  ]
  {aipproc}
\layoutstyle{6x9}

\begin{document}

\title{ Parametrization and Classification of 20 Billion LSST Objects: Lessons from SDSS }


\author{\v{Z}. Ivezi\'{c}}{
   address={University of Washington, Dept. of Astronomy, Box 351580, Seattle, WA 98195}
}

\author{T. Axelrod}{
  address={LSST Corporation, 933 N. Cherry Avenue, Tucson, AZ  85721}
}

\author{A.C. Becker}{
   address={University of Washington, Dept. of Astronomy, Box 351580, Seattle, WA 98195}
}

\author{J. Becla}{
  address={Stanford Linear Accelerator Center, Stanford University, Stanford, CA 94309}
}

\author{K. Borne}{
  address={Dept. of Comput. and Data Sciences, George Mason
                University, MS 6A2, Fairfax, VA, 22030, USA}
}

\author{D.L. Burke}{
  address={Stanford Linear Accelerator Center, Stanford University, Stanford, CA 94309}
}

\author{C.F. Claver}{
  address={National Optical Astronomy Observatory, 950 N. Cherry Ave, Tucson, AZ 85719}
}

\author{K.H. Cook}{
  address={Lawrence Livermore National Laboratory, 7000 East Avenue,
    Livermore, CA 94550}
}

\author{A. Connolly}{
  address={University of Washington, Dept. of Astronomy, Box 351580, Seattle, WA 98195}
}

\author{D.K. Gilmore}{
  address={Stanford Linear Accelerator Center, Stanford University, Stanford, CA 94309}
}

\author{R.L. Jones}{
  address={University of Washington, Dept. of Astronomy, Box 351580, Seattle, WA 98195}
}

\author{M. Juri\'{c}}{
  address={Institute for Advanced Study, 1 Einstein Drive, Princeton, NJ 08540}
}

\author{S.M. Kahn}{
  address={Stanford Linear Accelerator Center, Stanford University, Stanford, CA 94309}
}

\author{K-T. Lim}{
  address={Stanford Linear Accelerator Center, Stanford University, Stanford, CA 94309}
}

\author{R.H. Lupton}{
  address={Department of Astrophysical Sciences, Princeton University, Princeton, NJ 08544}
}

\author{D.G. Monet}{
  address={U.S. Naval Observatory Flagstaff Station, 10391 Naval Observatory
              Road, Flagstaff, AZ 86001}
}

\author{P.A. Pinto}{
 address={Steward Observatory, The University of Arizona, 933 N Cherry Ave., Tucson, AZ 85721}
}

\author{B. Sesar}{
 address={University of Washington, Dept. of Astronomy, Box 351580, Seattle, WA 98195}
}

\author{C.W. Stubbs}{
  address={Center for Astrophysics, Harvard University, 60 Garden St., Cambridge, MA 02138 }
}

\author{J. A. Tyson}{
 address={Physics Department, University of California, One Shields Avenue, Davis, CA 95616}
}

\begin{abstract}
The Large Synoptic Survey Telescope (LSST) will be a large, wide-field 
ground-based system designed to obtain, starting in 2015, multiple images 
of the sky that is visible from Cerro Pachon in Northern Chile. About 
90\% of the observing time will be devoted to a deep-wide-fast survey mode 
which will observe a 20,000 deg$^2$ region about 1000 times during the 
anticipated 10 years of operations (distributed over six bands, $ugrizy$). 
Each 30-second long visit will deliver 5$\sigma$ depth for point sources 
of $r\sim24.5$ on average. The co-added map will be about 3 magnitudes 
deeper, and will include 10 billion galaxies and a similar number of stars. 
We discuss various measurements that will be automatically performed for 
these 20 billion sources, and how they can be used for classification and 
determination of source physical and other properties. We provide a few 
classification examples based on SDSS data, such as color classification of 
stars, color-spatial proximity search for wide-angle binary stars, 
orbital-color classification of asteroid families, and the recognition of 
main Galaxy components based on the distribution of stars in the 
position-metallicity-kinematics space. Guided by these examples, we anticipate 
that two grand classification challenges for LSST will be 1) rapid and robust 
classification of sources detected in difference images, and 
2) {\it simultaneous} treatment of diverse astrometric and photometric time 
series measurements for an unprecedentedly large number of objects. 
\end{abstract}
\maketitle


\section{Parametrization of LSST Objects}

Instead of discussing numerous classification techniques that are the main 
topic of this conference, we focus on the properties of measured parameters 
that represent input to various classification algorithms and methods. 
We summarize the anticipated measurements for LSST, and provide a few 
classification examples based on similar SDSS data.

\subsection{A brief overview of LSST}

The Large Synoptic Survey Telescope (LSST) will be a large, wide-field 
ground-based system designed to obtain multiple images covering the sky 
that is visible from Cerro Pachon in Northern Chile. The system design is
driven by four main science themes: probing dark energy and dark matter, 
taking an inventory of the Solar System, exploring the transient optical 
sky, and mapping the Milky Way. For a detailed discussion of LSST science 
drivers, baseline design, and anticipated data products, please see 
Ivezi\'{c} et al. (2007a). 

Briefly, the current baseline 
design, with an 8.4m (6.5m effective) primary mirror, a 9.6 sq.deg. field 
of view, and a 3.2 Gigapixel camera, will allow about 10,000 deg$^2$ 
of sky to be covered each night (using pairs of 15-second exposures), with
an average revisit time of three nights and 5-sigma depth for point sources 
of $r\sim24.5$. The survey area will include 30,000 deg$^2$ with
$\delta<34.5^\circ$, and will be imaged multiple times in six bands, $ugrizy$, 
covering the wavelength range 320--1050 nm. About 90\% of the observing time 
will be devoted to a deep-wide-fast survey mode which will observe, starting 
in 2015, a 20,000 deg$^2$ region about 1000 times during the anticipated 
10 years of operations (including all six bands). These data will result in 
databases including 10 billion galaxies and a similar number of stars, and 
will serve the majority of science programs.

\subsection{The measured parameters for LSST Sources and Objects}

The rapid cadence of the LSST observing will produce an enormous
volume of data, $\sim$30 TB per night, leading to a total database over 
the ten years of operations of 60 PB for the compressed raw data, and 30 PB for the 
catalog database. The total data volume after processing will be several 
hundred PB, processed using $\sim$100 TFlops of computing power. Processing 
such a large volume of data, converting the raw images into a faithful
representation of the Universe, automated data quality assessment, and 
archiving the results in useful form for a broad community of users is a 
major challenge.

The two main data deliverables will be transient event reporting, and yearly 
data releases. The transient event reporting system 
will send out alerts to the community within 60 seconds of
completing the image readout. Yearly data releases are intended 
to produce the most completely analyzed data products of the survey (in
particular those that measure very faint objects and cover long time
scales). The transient event reporting system will be based on difference 
image analysis and will include classification based on measured 
properties in the event image, as well as on properties of the local
background determined from template images (see Bailey et al. 2007,
Becker 2008, and Borne 2008 for examples of ongoing work). We limit
discussion here to recurrent sources (as opposed to one-time events
such as various explosions). 

For each of $\sim$20 billion objects detected in deep images, 
the time series of positions, $\alpha(t)$ and $\delta(t)$, and magnitudes,
$m(t); m=(u,g,r,i,z,y)$, will be measured (detections in individual
images are called ``sources'', and they are associated into ``objects''). 
The systematic errors in positions will be of the order 10 mas, and
photometric errors will be about 5 millimag at the bright end. 
While automated pipelines will measure several types of magnitudes (e.g. point 
spread function, model, Petrosian, etc.), as well as higher image moments, 
such as image shape and size, for simplicity here we only consider the above 
listed generic quantities (however, for a spectacular example of an astronomical 
source with time-varying morphology see Rest et al. 2005). Morphological information 
will be used for star-galaxy separation. We ignore this important classification
problem in this paper. Of course, morphological information will be
used to separate comets from asteroids as well, to classify galaxies (e.g. via 
bulge-disk decomposition), and the measurements of galaxy shapes
will form the basis of weak lensing studies.
 
Given the above three time series, $\alpha(t), \delta(t)$, and $m(t)$,
sampled about 1000 times over 10 years, for a sample of 20 billion objects,
it is not obvious how to perform efficient and robust classification.
Even if the measurement errors are ignored, there are about 60 trillion 
numerical values to analyze! Alternatively, one needs to quantitatively
describe a 3000-dimensional space populated by 20 billion points. 
The current practice and experience is such that one would not blindly 
feed this dataset to a computer algorithm and hope for an ultimate
and complete classification. Rather, both the choice of measurements
and applied computational techniques would be driven by specific science 
goals. For example, 
\begin{enumerate}
\item For slow moving objects such as stars (motion is typically 
smaller than arcsec per year), $\alpha(t)$ and $\delta(t)$ are modeled as 
a superposition of linear motion due to star's space velocity, the reflex 
motion due to Earth's revolution (trigonometric parallax), and possibly of
the orbital motion in a multiple star system. Displacements between two
successive observations are much smaller than the seeing disk, and thus
there is practically no ambiguity when associating different detections. 
This modeling yields trigonometric parallax, proper motion vector, and
orbital properties. 
\item For fast moving objects such as asteroids and other solar system 
objects (motion is typically larger than arcmin per year, and for nearby
asteroids faster than degree per day), displacements between two successive 
observations are much larger than 
the seeing disk, and often can be larger than the typical distance between
two (stationary) objects. Thus, associations of different detections may be 
highly uncertain, and to resolve the ambiguity $\alpha(t)$ and $\delta(t)$ 
are required to be consistent with a nearly-Keplerian orbital motion. 
This analysis yields six orbital parameters for each object.
\item The time-averaged values of $m(t)$ enable the construction of 
color-magnitude and color-color diagrams. These diagrams can be used to 
classify objects, as well as to derive additional, typically more physically
relevant, parameters, such as photometric redshift for galaxies and
quasars, photometric distance, effective temperature and metallicity 
for stars, and taxonomy for asteroids.
\item 
Detailed properties of light curves, $m(t)$, such as shape, 
amplitude and period for periodic variables, or the structure function
for aperiodic variability, often enable powerful classification 
schemes (e.g. the classification of variable stars in the amplitude-period 
diagram). The current LSST plans call for automatic computation of 
numerous parameters that are designed to capture detailed statistical 
properties of light curves (e.g. low-order moments, period, amplitude).
\item Measurements for individual objects need not be analyzed in isolation;
rather, they can be considered in relation to measurements of neighboring 
objects. 
\end{enumerate}

We proceed to provide concrete examples of each type of classification 
based on SDSS data. 

\section{Examples of Classification}

In many aspects, SDSS data are similar to anticipated LSST data.
In particular, the photometric systems are similar, and SDSS provides 
some time domain data (light curves, proper motion, solar system objects).
LSST will extend the coverage down to 60 sec timescale, and to much fainter objects.
Thus, the ongoing classification work based on SDSS data can serve as 
a preview of what will be possible to do with LSST on a much larger scale.

\begin{figure}
\includegraphics[height=.26\textheight]{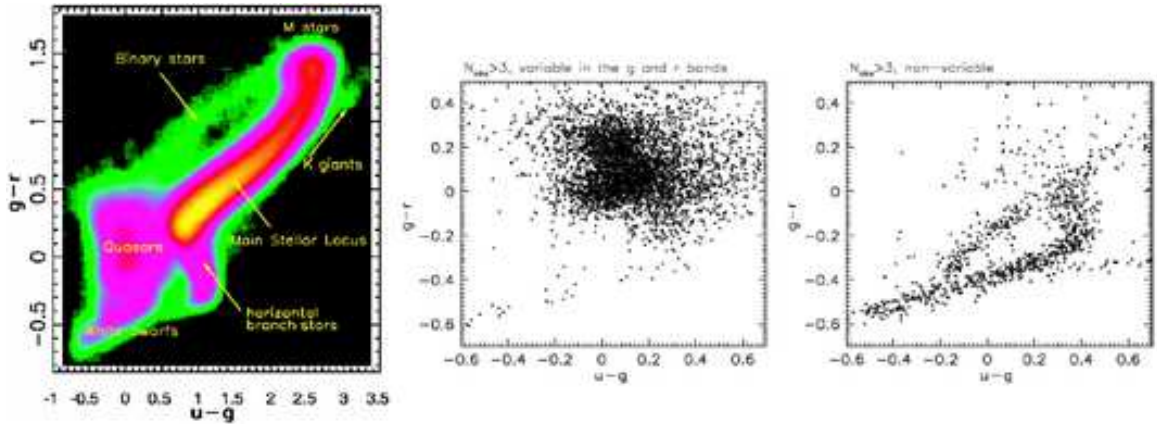}
\caption{
The left panel shows the $g-r$ vs. $u-g$ color-color diagram for two million
stars measured by SDSS (adapted from Smol\v{c}i\'{c} et al. 2004). Accurate
$ugr$ photometry enables detailed and robust classification. The middle 
and right panel show the zoomed-in lower-left corner of this diagram (adapted 
from Ivezi\'{c} et al. 2007b), with sources classified by their photometric 
variability into highly variable (middle; dominated by quasars) and 
non-variable (right; dominated by H and He white dwarfs). LSST will obtain 
such measurements and enable resulting classification for billions of stars 
and millions of quasars.
\label{fig:ugr}}
\end{figure}

\subsection{    Classification based on colors    } 
\begin{figure}
\includegraphics[height=.25\textheight]{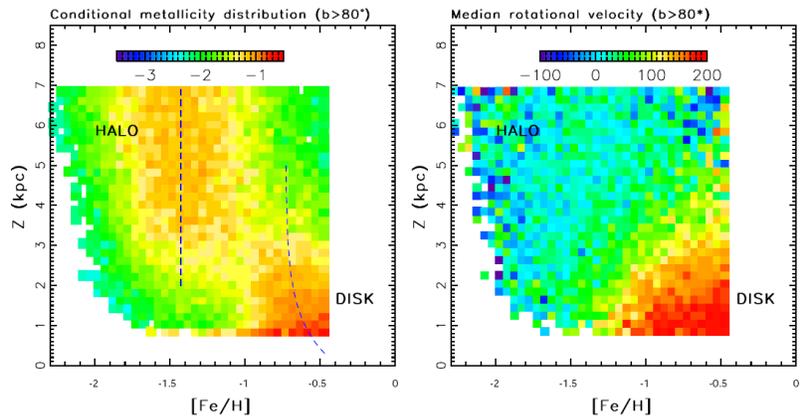}
\caption{
An illustration of stellar classification based on $ugr$ photometry and 
proper motion measurement. The left panel shows the photometric metallicity 
distribution, as a function of the distance from the Galactic plane, for 
about 60,000 F/G main-sequence stars within 10 degrees from the North 
Galactic Pole (adapted from Ivezi\'{c} et al. 2008). The distribution is 
displayed on a logarithmic scale and color-coded as shown in the inset. 
Two distinct Galaxy components, halo and disk, are marked. The dashed 
lines mark the median metallicity for each component. These two components 
with distinct metallicity distributions also have different kinematics, 
as illustrated in the right panel. The color-coded map shows the median 
rotational velocity component for the same stars as in the left panel. 
The velocity is determined from displacements of stars on the sky over 
half a century that lapsed between the Palomar Observatory Sky Survey in the 
1950s and SDSS. The high-metallicity disk stars have large rotational 
velocity (about 200 km/s, see the inset), while the low-metallicity halo 
stars display behavior consistent with no net rotation. 
\label{fig:pm}
}
\end{figure}

The position of a source in the $g-r$ vs. $u-g$ color-color diagram 
(Fig.~\ref{fig:ugr}) contains rich information that can be used for 
preliminary source classification.  Determination of detailed stellar 
properties such as effective temperature and metallicity can further
improve the resulting classification (see left panel in Fig.~\ref{fig:pm}), 
as well as the addition of data from more bandpasses. 

\begin{figure}
  \includegraphics[height=.20\textheight]{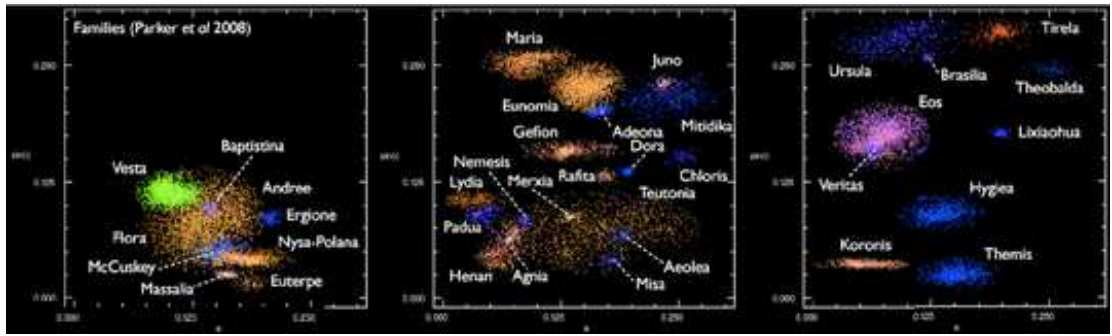}
\caption{
An illustration of classification based on photometry and orbital
motion parameters. The three panels show the distribution of main-belt 
asteroids in the orbital inclination vs. eccentricity space, for three
regions selected by the orbital semi-major axis (adapted from 
Parker et al. 2008). Each dot represents one asteroid, and is
color-coded according to its colors measured by SDSS. The families
are defined by simultaneously applying Gaussian filters in orbital 
and color space.   
\label{fig:asteroids}
}
\end{figure}

\subsection{    Classification based on motion  } 

The moving objects naturally split into two classes: slow stellar
motion and fast motion of solar system objects. 
Fig.~\ref{fig:pm} illustrates how proper motion measurements can be 
used to classify stars into disk and halo populations. 
This can be applied to other types of objects.  The simultaneous 
use of orbital parameters and color to associate asteroids into 
families (remnants of larger bodies destroyed in collisions) is 
illustrated in Fig~\ref{fig:asteroids}.

\subsection{    Classification based on photometric variability  } 

The color information can be augmented with variability information to 
further improve source classification. For example, the right panel in 
Fig.~\ref{fig:ugr} shows that the $u-g$ color distribution of {\it non-variable} 
sources with blue $g-r$ colors is tri-modal -- evidently, the information 
content is very high! The bluest branch is consistent with He white dwarfs, 
the middle branch with hydrogen white dwarfs, and the reddest branch is 
made of blue horizontal branch stars (see Ivezi\'{c} et al. 2007b for more 
details and references). 

\begin{figure}
  \includegraphics[height=0.20\textheight]{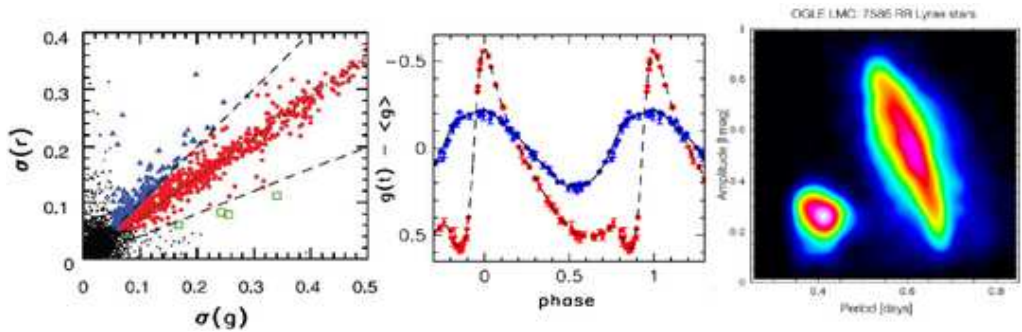}
\caption{
An illustration of classification based on colors and light
curve shape, period and amplitude. The left panel shows a 
relationship between the photometric root-mean-square scatter 
measured in SDSS $g$ and $r$ bands for point sources with colors 
typical of RR Lyrae stars (marked by ``horizontal branch stars''
in Fig.~\ref{fig:ugr}). The two dashed lines mark the
region consistent with RR Lyrae light curves (for details
see Sesar et al. 2007). The middle panel shows examples
of RR Lyrae light curves (blue: c type, red: ab type)
measured by SDSS. The right panel shows the distribution 
of RR Lyrae stars of both types in the amplitude vs. period
diagram (measured by the OGLE survey; adapted from 
Eyer \& Mowlavi 2007). 
\label{fig:RRLyr}
}
\end{figure}

The statistical properties of light curves have long been used for 
robust classification, especially when multi-color data are available.
Fig.~\ref{fig:RRLyr} illustrates classification of RR Lyrae stars 
with the aid of multi-color light curves. It is estimated that 
LSST will detect and parametrize about 100 million variable stars.

\subsection{    Classification involving neighbors  } 

\begin{figure}
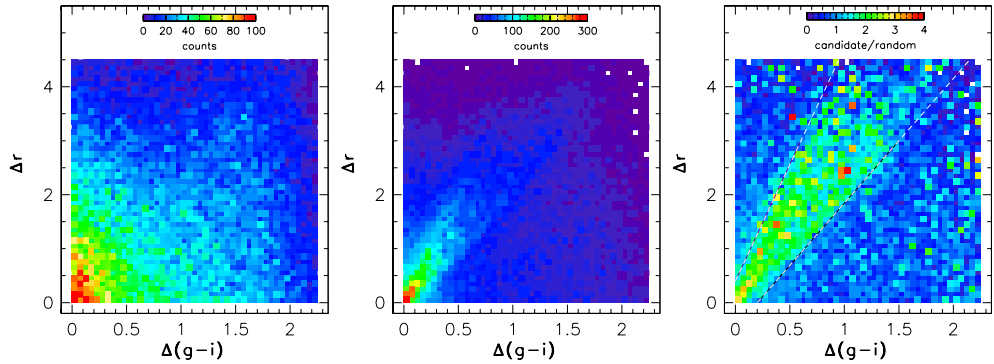

  \includegraphics[height=.22\textheight]{ivezic_fig5a}
  \includegraphics[height=.22\textheight]{ivezic_fig5b}
  \includegraphics[height=.22\textheight]{ivezic_fig5c}
\caption{An illustration of the simultaneous use of photometry and 
astrometric proximity to nearby sources to select pairs of stars 
that are good candidates for wide-angle binary systems (adapted from 
Sesar et al. 2008). The left panel shows the distribution of stellar 
pairs from a random SDSS sample in the magnitude difference vs. color 
difference diagram. The middle panel shows analogous diagram for 
stellar pairs with component separation in the range 3--4 arcsec. 
This sample shows an excess of pairs whose magnitude and color differences 
are consistent with the same distance from us (via photometric
parallax relation), as demonstrated by the ratio of the two maps
shown in the right panel. 
\label{fig:wab} 
}
\end{figure}

In addition to measurements for an individual object, measurements
of its neighboring objects (either in spatial proximity, or using an 
arbitrary parameter) can be used in classification. For example, Sesar 
et al. (2008) developed a method for selecting candidate wide-angle 
stellar binary systems by requiring i) spatial proximity of two stars 
on the sky, ii) similarity of their photometric distances, and iii) 
similarity of their proper motion vectors (Fig.~\ref{fig:wab}). 
Strong gravitational lensing of supernovae is another example.

\section{   Discussion  } 

The classification challenge for recurrent LSST sources can be 
phrased as: {\it given three time series, $\alpha(t), \delta(t)$, 
$m(t)$, and associated morphology, sampled 1000 times over 10 years, for 20 billion objects,
assign objects to known populations, discover new populations,
and quantify their behavior}. The current practice and experience 
argue that there is no ``one size fits all'' solution; rather,
the choice of measurements and applied computational techniques 
is driven by specific science goals, as briefly illustrated by 
examples presented here. The requirements for rapid and robust
classification of transient sources present additional challenges.
We invite experts in the fields of parametrization and classification
to join LSST effort and contribute timely solutions to these
challenges.


\begin{theacknowledgments}
We thank our numerous SDSS (www.sdss.org) and LSST (www.lsst.org)
collaborators for their valuable contributions and helpful discussions.
\end{theacknowledgments}

\bibliographystyle{aipproc}   


\IfFileExists{\jobname.bbl}{}
 {\typeout{}
  \typeout{******************************************}
  \typeout{** Please run "bibtex \jobname" to optain}
  \typeout{** the bibliography and then re-run LaTeX}
  \typeout{** twice to fix the references!}
  \typeout{******************************************}
  \typeout{}
 }


\end{document}